\documentclass[useAMS,usenatbib]{mn2e}
\usepackage{times}
\usepackage{graphicx}

\newcommand{\kms}   {km~s$^{-1}$}

\newcommand{\eg}    {e.\,g.}
\newcommand{\ie}    {i.\,e.}

\newcommand{\sii}   {[S\,{\sc ii}]}
\newcommand{\nii}   {[N\,{\sc ii}]}
\newcommand{\nin}   {[N\,{\sc i}]}
\newcommand{\oiii}   {[O\,{\sc iii}]}

\newcommand{\vlsr}  {$V_\rmn{LSR}$}

\title[Integral Field Spectroscopy of HH 262 ]{Integral Field 
Spectroscopy of HH 262: The Spectral Atlas}

\author[R. L\'opez et al.]
{R. L\'opez,$^{1}$\thanks{E-mail:
\mbox{rosario@am.ub.es;
bgarcia@iac.es };
sanchez@caha.es;
\mbox{gabriel.gomez@gtc.iac.es;
robert.estalella@am.ub.es;
angels.riera@upc.edu}
}
B. Garc\'\i a-Lorenzo $^{2}$\footnotemark[1]
S. F. S\'anchez,$^{3}$\footnotemark[1] 
G. G\'omez,$^{4,2}$\footnotemark[1]
R. Estalella,$^{1}$\footnotemark[1]
 and
\newauthor 
A. Riera$^{5}$\footnotemark[1] 
\thanks{Based on observations collected at the Centro Astron\'omico Hispano
Alem\'an (CAHA) at Calar Alto, operated jointly by the Max-Planck Institut f\"ur
Astronomie and the Instituto de Astrof\'{\i}sica de Andaluc\'{\i}a (CSIC) 
}\\
$^{1}$Departament d'Astronomia i Meteorologia, Universitat de Barcelona,
Mart\'{\i} i Franqu\`es 1, E-08028 Barcelona, Spain\\
$^{2}$Instituto de Astrof\'{\i}sica de Canarias, E-38200 La Laguna, Spain\\
$^{3}$Centro Astron\'omico Hispano-Alem\'an de Calar Alto, Jes\'us Durban Remon
2-2, E-04004 Almer\'{\i}a, Spain.\\
$^{4}$GTC Project Office, GRANTECAN S.A. (CALP), E-38712 Bre\~na Baja, La Palma,
Spain.\\
$^{5}$Dept.\ F\'{\i}sica i Enginyeria Nuclear. EUETI de Barcelona.
Universitat Polit\`ecnica de Catalunya. Compte d'Urgell 187, E-08036 Barcelona,
Spain
}

\begin{document}

\date{Accepted . Received ; }
\pagerange{\pageref{firstpage}--\pageref{lastpage}} \pubyear{2008}

\maketitle

\label{firstpage}

\begin{abstract}

HH~262 is a group of emitting knots displaying an ''hour-glass'' morphology in the
H$\alpha$ and \sii\ lines, located 3.5 arcmin to the northeast of the young
stellar object  L1551-IRS5, in Taurus. We present new results of the kinematics
and physical conditions of HH~262 based on  Integral Field Spectroscopy covering
a field of $\sim$ 1.5 $\times$ 3 arcmin$^2$, which includes all the bright knots
in HH~262. These data show  complex kinematics and  significant variations  in 
physical conditions  over the mapped region of HH~262 on a  spatial scale of
$\leq$ 3 arcsec. A new result derived from the IFS data is  the weakness of the
\nii\ emission (below detection limit in most of the mapped region of HH~262),
including the brightest central knots. Our data  reinforce the association of 
HH~262  with the redshifted lobe of the evolved molecular outflow L1551-IRS5. 
The interaction of this outflow with a younger one, powered by L1551 NE, around
the position of HH~262 could give rise  to the complex morphology and
kinematics of HH~262. 

\end{abstract}


\section{Introduction}

HH~262 is the designation of a group of emitting knots, surrounded by  more
diffuse emission, located  $\sim$ 3.5 arcmin to the  northeast of the source 
L1551-IRS5  in Taurus. The two brightest HH~262 condensations (named GH9 and
GH10) were first reported by \citet{Rod89} and \citet{Gra90}. In the
H$\alpha$+\nii\ and \sii\ narrow-band images, HH~262 shows a rather chaotic,
''hour-glass''-shaped morphology that is quite different from the collimated
structures found in most of the Herbig--Haro (HH) jets (\citealt{Gar92},  
\citealt{Lop95}). Apparently, the morphology of HH~262 seems to  trace a
wind-blown cavity from an embedded source located close to the GH10 knot.
However, no  point source  suitable for powering the HH~262  emission has been
currently reported at this location.  Since its discovery, the origin of HH~262,
as well as its relationship  with other outflows in the L1551 star-forming
region, has been  matter of discussion. \citet{Rod89} first suggested that the
GH9 and GH10 knots were associated  with the redshifted lobe of the  molecular
outflow powered by the source L1551-IRS5. They proposed that GH9 and GH10 may be
the redshifted counterpart of a chain of HH bright knots  (which includes the
well-known HH objects 28 and 29) extending from L1551-IRS5 to the southwest up
to $\sim$ 5 arcmin, and suggested that  all these optical knots  traced the
walls of a wind cavity evacuated by the CO outflow powered by L1551-IRS5.

\begin{figure}
\centering
\includegraphics[width=\hsize]{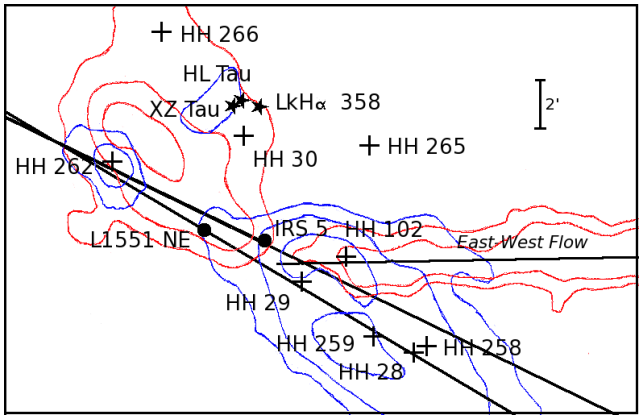}
\caption{
Diagram of the HH~262 region showing the location of HHs and  YSOs. 
The contours correspond to the blueshifted and redshifted CO
emission from the three molecular outflows identified in the region
(two outflows in the NE-SW direction, associated with the YSOs L1551
IRS 5 and L1551 NE, and the east-west outflow; \citealt{Mor06}). 
The straight lines are drawn along the outflow axes.
\label{diagram1}}
\end{figure}

Later on, \citet{Gra92} obtained a long-slit spectrum across the GH9 and  GH10
knots. From the H$\alpha$ line, they derived redshifted radial velocities
covering a range extending over $\sim$ 100~\kms  and measure peak  velocities of +40~\kms\ 
and +90~\kms\ and average velocities of +20~\kms and +45~\kms in the GH9 and
GH10 knots, respectively.  These results reinforced the suggestion that GH9 and
GH10 were associated with the redshifted lobe of the CO outflow powered by
L1551-IRS5.  The kinematics of HH~262, obtained by \citet{Lop98} from proper
motion measurements of the brightest knots, and from radial velocities derived
from Fabry-Perot observations in the \sii\ $\lambda$ 6716 \AA\ line, is also
consistent with the interpretation of the HH~262 origin mentioned before.
However, \citet{Dev99} were uncertain whether the exciting source of HH~262 is
L1551-IRS5 or  L1551 NE, another younger, embedded proto-binary source located
$\sim$ 150 arcsec northeast of IRS5 (\citealt{Eme84}, \citealt{Rod95}).  L1551
NE also drives a molecular CO outflow, so that it is likely that the outflows
powered by these two sources are interacting.  Further observations from
\citet{Sto06} showed evidence of blueshifted molecular gas arising close to
L1551 NE that  extends to the northeast of the source and encompasses HH~262. As
pointed out by \citet{Dev99} and \citet{Mor06}, HH~262 could be related to
either the L1551-IRS5 or L1551 NE outflows, its rather chaotic  morphology 
originating through the interaction  between  the  molecular outflows powered by
these two sources (see diagram of Fig.\ref{diagram1}).  More recent sub-mm continuum observations at 850 $\mu$m by
\citet{Mor06}  detected emission around HH~262 that were interpreted as produced
by warm dust entrained by the molecular outflows. HH~262 is thus a suitable
place to search for  optical  signatures of the interactions among supersonic
outflows. In addition, because of its morphology, HH~262 is better suited for  
performing Integral Field Spectroscopy (IFS) than long-slit observations when 
the kinematics and physical conditions of the  emission need to be analysed.

With the aim of advancing in the study of HH~262, we included this target within
a program of Integral Field Spectroscopy of Herbig--Haro objects using the 
Potsdam Multi-Aperture Spectrophotometer (PMAS) in the wide-field IFU mode
PPAK.  We present in this paper new results on the kinematics and physical
conditions of HH~262 obtained from this IFS observing program. The paper is
organized as follows. The observations and data  reduction are described in \S\
2.  Results are given in \S\ 3: The integrated maps in \S\ 3.1, the channel maps
in \S 3.2, and the spectral atlas of HH~262, including discussions of the
characteristic of the more relevant knots in \S\ 3.3. A global discussion and
the main conclusions are given  \S\ 4. 

\section[]{Observations and Data Reduction}

Observations of HH~262 were made on 22 November 2004 with the 3.5-m  telescope
of the Calar Alto Observatory (CAHA). Data were acquired with the  Integral
Field Instrument Potsdam Multi-Aperture Spectrophotometer PMAS \citep{Rot05}
using the PPAK configuration that has 331 science fibres, covering an hexagonal
FOV of 74$\times$65 arcsec$^2$ with a spatial sampling of 2.7 arcsec per fibre,
and 36 additional fibres to sample the sky (see Fig.\ 5 in \citealt{Ke06}). The
I1200 grating was used, giving an effective sampling  of 0.3 \AA\ pix$^{-1}$ 
($\sim15$  km~s$^{-1}$ for H$\alpha$) and covering the wavelength range
$\sim6500$--7000 \AA, thus including characteristic HH emission lines in this
wavelength range (H$\alpha$, \nii\ $\lambda\lambda$6548, 6584 \AA\  and \sii\ 
$\lambda\lambda$6716, 6731 \AA). The spectral resolution (\ie\ instrumental
profile) is $\sim2$ \AA\  FWHM ($\sim90$ km~s$^{-1}$) and the accuracy in the
determination of the position of the line centroid is $\sim0.2$ \AA\  ($\sim10$
km~s$^{-1}$ for the  strong observed emission lines). Eight overlapped pointings
of 1800-s exposure time each were observed to obtain a mosaic of $\sim$ 1.5
$\times$ 3 arcmin$^2$  to cover almost the entire emission from HH~262.  Note,
however, that  this mosaic does not cover the extended, arc--shaped emission
known  as HH~262E, which is most probably related  to HH~262.  In Table 1 we
list the centre positions of each  pointing and the corresponding HH~262 knot
identification following the nomenclature of \citet{Gar92},  extended by
\citet{Lop95}.

Data reduction was performed using a preliminary version of the {\small R3D}
software \citep{Sa06}, in combination with {\small IRAF}% 
\footnote{{\small
IRAF} is distributed by the National Optical Astronomy Observatories, which are
operated by the Association of Universities for Research in Astronomy, Inc.,
under cooperative agreement with the National Science Foundation.} and the
{\small Euro3D} packages \citep{Sa04}. The reduction consists of the standard
steps for fibre-based integral field spectroscopy. A master bias frame was
created by averaging all the bias frames observed during the night and
subtracted from the science frames. The location of the spectra in the CCD was
determined using a continuum-illuminated exposure taken before the science
exposures. Each spectrum was extracted from the science frames by co-adding the
flux  within an aperture of 5 pixels along the cross-dispersion axis for each
pixel in the dispersion axis, and stored in a row-stacked-spectrum (RSS) file
\citep{Sa04}. At the date of the observations we lacked of a proper calibration
unit for PPAK. Therefore the wavelength calibration was performed iteratively,
following all the steps described in \citet{Lop08} for HHL~73,  another target
of our HH  IFS survey.  The final wavelength solution was set by using the sky
emission lines found in the observed wavelength range. The  accuracy achieved
for the wavelength calibration was better than  $\sim0.1$  \AA\ ($\sim5$
km~s$^{-1}$). Observations of a standard star were used to perform a relative
flux calibration. A final datacube  containing the 2D spatial plus the spectral
information of HH~262 was then created from the 3D data by using {\small Euro3D}
tasks  to interpolate the data spatially until reaching a final grid of 2.16
arcsec of spatial sampling and with a spectral sampling of 0.3 \AA\ . Further
manipulation of this datacube, devoted to obtain  integrated emission line maps,
channel maps and position--velocity maps, were made using several common-users
tasks of {\small STARLINK}, {\small IRAF} and {\small GILDAS} astronomical
packages and {\small IDA} \citep{Ga02}, an specific {\small IDL} software to
analyse 3D data.

\begin{table}
\begin{minipage}{130mm}
\caption{Positions of the single pointings of the HH~262 mosaic.
\label {tjcmt}}
\begin{tabular}{@{}lcc@{}}
\hline
\multicolumn{2}{c}{Position} 
&knot\\
\hline
04$^{\rm h}$31$^{\rm m}$58$\fs7$&+18$\degr$12$\arcmin$07$\arcsec$& GH9\\
04$^{\rm h}$32$^{\rm m}$01$\fs1$&+18$\degr$12$\arcmin$35$\arcsec$& K18\\
04$^{\rm h}$32$^{\rm m}$00$\fs5$&+18$\degr$12$\arcmin$05$\arcsec$& K19\\
04$^{\rm h}$32$^{\rm m}$00$\fs9$&+18$\degr$11$\arcmin$53$\arcsec$& K20\\
04$^{\rm h}$32$^{\rm m}$00$\fs1$&+18$\degr$11$\arcmin$33$\arcsec$& GH10W\\
04$^{\rm h}$32$^{\rm m}$01$\fs1$&+18$\degr$11$\arcmin$24$\arcsec$& GH10E\\
04$^{\rm h}$32$^{\rm m}$00$\fs0$&+18$\degr$11$\arcmin$17$\arcsec$& K26\\
04$^{\rm h}$31$^{\rm m}$59$\fs7$&+18$\degr$10$\arcmin$30$\arcsec$& K22\\
\hline
\end{tabular}
\end{minipage}
\end{table}

\section{Results}

\subsection{Integrated line-emission maps}

\subsubsection{Emission line images}

\begin{figure}
\centering
\includegraphics[width=\hsize]{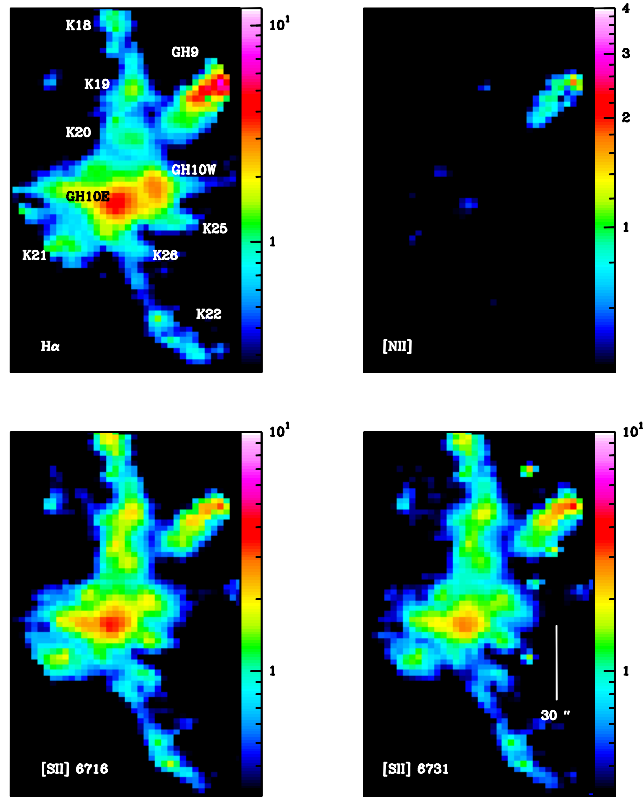}
\caption{
Integrated maps of the HH~262 emission obtained from the datacube 
by
integrating the signal within the wavelength range including the 
line labelled in each panel:
(H$\alpha$ : $\lambda$6558--6569 \AA, 
\nii\ :	    $\lambda$6580--6586 \AA,  
\sii\ : $\lambda$6716  6712--6719 \AA\  and 
\sii\ : $\lambda$6731  6728--6734 \AA ).  
Fluxes have been displayed in a logarithmic scale, in  units of 10$^{-16}$
erg~s$^{-1}$~cm$^{-2}$. In all the
maps, north is up and East is to the left. 
The HH~262 knots have been labelled in
the H$\alpha$ panel
\label{flux}}
\end{figure}

Emission from
H$\alpha$ and \sii\ $\lambda\lambda$6716, 6731 \AA\  was detected in
all the mapped region. In contrast, emission from the \nii\
$\lambda\lambda$6548, 6584 \AA\ lines  was not detected, except in the 
knot GH9,
towards the northwestern side of HH~262.  Narrow-band images for each of the
lines were generated from the datacube.  For each position, the flux of the
line was obtained by integrating the signal over the whole wavelength  range
covered by the line  and subtracting the continuum
emission obtained from the wavelengths adjacent to the 
line. The
narrow-band maps of HH~262 obtained in this way are shown in Fig.\ref{flux}

As  can be seen from the panels of the figure, the H$\alpha$ and \sii\ emissions
reveal a similar morphology, although the emission of  some the knots is 
stronger  in
\sii\ than in  H$\alpha$, while  the brightness of the \sii\ and H$\alpha$
emission  is similar for the other knots, in good agreement with
the results found in previous CCD narrow-band images (\citealt{Gar92}). 

Our IFS data allow us to obtain the first \nii\ narrow-band image of HH~262.
Because there are no CCD images acquired through a line filter with a passband
narrow enough to isolate the \nii\ lines from H$\alpha$, the  morphology of
HH~262 in the  \nii\ lines has remained unknown up to now.  Note that the 
morphology of HH~262 in  \nii\  is  different than in  H$\alpha$ and \sii\ :
\ie\ HH~262 does not show its characteristic ''hour-glass'' structure because the
emission from \nii\ was detected only in knot GH9. In the other knots, \nii\
emission was only marginally detected or not detected at all.

\subsubsection{Kinematics}

\begin{figure}
\centering

\includegraphics[angle=-90,width=\hsize]{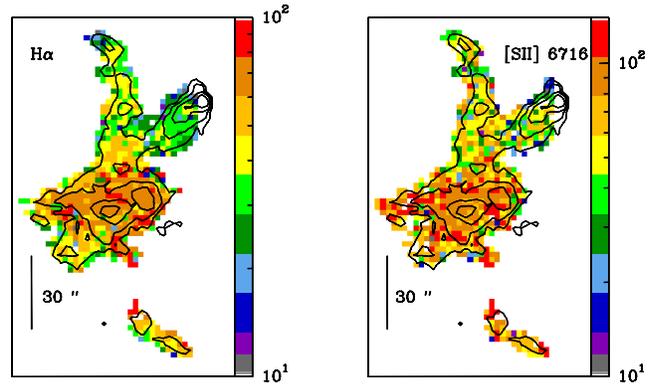}
\caption{
Velocity maps of the line centroid. Velocities (km~s$^{-1}$) are referred to the
\vlsr\ of the parent cloud. Contours of the integrated flux of the
H$\alpha$ line have been overlaid to help with the knots identification. 
\label{vel}}
\end{figure}

Radial velocity\footnote{All the velocities in the paper are referred to the
local standard of rest (LSR) frame. A   $\mbox{\vlsr}=+7.0$ \kms\ for the parent
cloud has been taken from \citet{Gra92}.} maps for the  H$\alpha$ and \sii\
emissions were obtained from the line centroids of a Gaussian fit to the
line emission. All the HH~262 knots show redshifted velocities in both lines
(H$\alpha$  and \sii), the velocity field being rather complex (Figure
\ref{vel}). The velocities derived for the H$\alpha$ emission range from $\sim
+25$ to $\sim +85$ km~s$^{-1}$ and, in general, are  $\sim 20$ km~s$^{-1}$ lower
than those derived from \sii\ emission,  which range from $\sim +40$ to $\sim
+110$ km~s$^{-1}$, the difference found between the H$\alpha$ and \sii\
velocities being reliable. The highest velocities are found towards the central
knots of HH~262 (a velocity  of $\sim +70$  in H$\alpha$ and and $\sim +100$
km~s$^{-1}$ in the  \sii\ $\lambda$ 6716 \AA\ line, respectively, is found
around GH10W).  The velocity decreases as we move  north and  south of  the
centre, with a steeper decrease   towards the north. The lowest  velocities
appear towards the  northwest, in the  GH9 knots, where a value of $\sim +30$
km~s$^{-1}$  for both lines is derived around its northern edge.  It is
interesting to note here that the velocities derived in this work  are in good
agreement with those found in previous studies (from long-slit spectroscopy of 
\citealt{Gra92}, and from  Fabry-Perot observations of \citealt{Lop98}).

The FWHMs of the H$\alpha$ and \sii\ lines, obtained from the Gaussian fit
and corrected from the instrumental profile, show the same behaviour: for both
emissions, the widest FWHM values are found towards the centre of HH~262 ($\sim
80$ km~s$^{-1}$ around the knot GH10E and $\sim 100$ km~s$^{-1}$ around GH10W),
the line widths diminishing when we move towards the north, reaching values 
of $\sim 30$ km~s$^{-1}$ around the knot 18. The FWHMs vary along GH9, ranging from
$\sim 40$ km~s$^{-1}$ (similar to those found around knots 19 and 20) in the
southern region of GH9, and increase towards the north (a FWHM of $\sim 65$
km~s$^{-1}$ is found around the GH9 NE edge).

\subsubsection{Physical conditions} 

Line-ratio maps of the integrated line emission were derived to explore the
spatial behaviour of the  electron density (from \sii\ 6716/6731) and the
excitation conditions (from \sii/H$\alpha$ and  from \nii/H$\alpha$ at the loci
where \nii\ emission was detected).

As a general trend, the electron density ($n_\rmn{e}$) derived in HH~262 is
low when compared with densities found in most of the HH jets.  
The highest $n_\rmn{e}$
values are found in GH9, coinciding  with the HH~262 loci where emission  from
\nii\ was detected ($n_\rmn{e}=490$--1000 cm$^{-3}$, increasing from the 
southeast
to the northwest along GH9). The electron density decreases as we move
towards the centre of HH~262, from north  ($n_\rmn{e}=270$ cm$^{-3}$ around K18)
to south ($n_\rmn{e}=100$ cm$^{-3}$ around K20). For the central knots (GH10E,
GH10W and K25),  the electron density lies close  to the low-density limit of
the \sii\ 6716/6731  line ratio sensitivity and thus only an upper limit for 
$n_\rmn{e}$ of 30--50 cm$^{-3}$ was estimated. The electron density 
increases slightly as we move from the centre  towards the south 
($n_\rmn{e}=120 $cm$^{-3}$ around K22) and southeast ($n_\rmn{e}=230$ cm$^{-3}$ around K21).
These values were derived from the \sii\ 6716/6731 ratio, using the  
{\small TEMDEN} task
of the {\small IRAF/STSDAS} package, and assuming $T_\rmn{e}=10^4$~K.

The derived spatial distribution of the \sii/H$\alpha$  line ratio  indicates
that  the excitation increases from east to west across HH~262. The lower line
ratio values (\sii/H$\alpha\simeq$0.6--0.9) are found towards the western region
of knots GH10W and GH9, reaching the lowest value (highest excitation)  around
GH9, coinciding with the loci  where \nii\ emission is detected. 
Furthermore, the trend found is  the \sii/H$\alpha$  line ratio decreases from
the northern K18--20 knots (with
\sii/H$\alpha\simeq$ 1.8--2.2) to the centre of HH~262 (GH10W-E knots, 
where \sii/H$\alpha\simeq$ 0.7--1.0). Beyond the
centre of HH~262, the \sii/H$\alpha$  line ratio increases slightly 
(the gas excitation decreases) as we move to the southern knots (K22, where 
\sii/H$\alpha\simeq$ 1.2).

As already mentioned, the \nii\ emission appears very weak and it was  detected
only marginally in the brightest HH~262 central knots  (see Fig.\ \ref{flux}).
To provide better confirmation of this result, we have  analysed an integrated
spectrum of each knot, obtained by  co-adding the signal of the spaxels inside
the aperture encircling the corresponding knot emission (see Section 3.3), thus
improving the SNR and the depth of the spectra. From these spectra,  we  have
derived  values for  the \nii\ flux emission (or an upper limit).  Table 2 lists
the flux of the \nii\ emission relative to the  rms of the spectrum,  the
\nii/H$\alpha$  line ratio and its estimated  error percentage. From the table,
we can see that  \nii/H$\alpha$ line ratios of $\sim$0.2 were obtained along GH9
knot. For the northern knots (K18, K19 and K20),  we estimated an upper limit of
$\sim$0.1 for this ratio (by considering the \nii\ flux emission enhancement at
a 3$\sigma$ rms level). For the central (GH10) and southern (K21, K26 and K22)
knots, \nii\ flux emission over the 3$\sigma$ rms  level was not detected at
all.

\begin{table}
\begin{minipage}{110mm}
\caption{Derived values  for the [NII] knots emission 
\label {tjcmt}}
\begin{tabular}{@{}llccc@{}}
\hline
\multicolumn{2}{c}{knot}
&[NII]/rms
&[NII]/H$\alpha$ 
&[NII]/H$\alpha$ error (\%)\\
\hline
K18&&$<$3.0&$<$0.10&\\
K19&&\phantom{<}3.0&\phantom{<}0.15 &10\\
K20&&\phantom{<}3.5&\phantom{<}0.18 &\phantom{1}8\\
&K20N&$<$3.0&$<$0.10 &\\
&K20S&\phantom{<}4.0&\phantom{<}0.18 &\phantom{1}7\\
GH9&&\phantom{<}3.5&\phantom{<}0.22 &\phantom{1}8\\
&GH9N-shell&\phantom{<}3.8&\phantom{<}0.20 &\phantom{1}8\\
&GH9W&\phantom{<}5.2&\phantom{<}0.17 &\phantom{1}5\\
&GH9E&\phantom{<}5.6&\phantom{<}0.20 &\phantom{1}5\\
GH10W&&$<$3.0&$<$0.05 &\\
GH10E&&$<$3.0&$<$0.08 &\\
K21&&$<$3.0&$<$0.10 &\\
K26&&$<$3.0&-\\
K22&&$<$3.0&-\\
\hline
\end{tabular}
\end{minipage}
\end{table}

We have searched the literature looking for HH spectra that would be similar to
those extracted for the HH~262 knots. Low \nii/H$\alpha$ line ratios ($\sim$
0.1--0.2), similar to those derived from the spectra of the HH~262 knots K19,
K20 and GH9, have been measured in the spectra of known HH jets (\eg\ HH~49,50:
\citealt{Sch80}; HH~111L: \citealt{Mor93}; HH~123: \citealt{Rei90}; HH~126B:
\citealt{Ogu91}; HH~34 (apex): \citealt{Mor93}). However, the spectra of these
jet knots,  covering a wider wavelength range to include other emission  lines,
seem to be quite different when compared with the HH~262 extracted spectra, and
would thus trace gas at different conditions. For example, the \sii/H$\alpha$ 
ratios in these jet knots are in general lower than the \sii/H$\alpha$ ratio
derived in  the HH~262 knots having similar \nii/H$\alpha$ values, and the
derived  $n_\rmn{e}$  of these jet knots are also higher than the densities
derived for HH~262.  Moreover, most of the above-mentioned knot spectra show
emission from \oiii\ lines.  Although the IFS data of HH~262  analysed here do
not cover a wavelength range that includes the \oiii\ lines, it should be
mentioned that we failed to detect \oiii\ emission at the location of HH~262  in
a CCD image of 3000-s exposure  time, acquired through an appropriate
narrow-band filter in a previous observing run. Thus, the HH~262 spectra appear
to have lower excitation than the  knots mentioned before having similar
\nii/H$\alpha$ line ratios. 

In order to compare the line ratios obtained in HH~262 with current models of
shock-ionization, it has to be noted that the HH~262 fluxes have not been
corrected for reddening. Thus the derived line ratios are somewhat affected by
differential extinction. However, since the wavelengths of the lines considered
are close, we expect that the differences between the observed and the corrected
line ratios are not critical for model comparison. For instance, for other HH
objects where corrected fluxes are available, the differences found are lower
than
1~\% and 15~\% for the \nii/H$\alpha$  and \sii/H$\alpha$ ratios
respectively (see \eg\ \citealt{Rag96} and references therein).

Line ratios such as those derived from the HH~262 spectra can be reproduced in a
framework of shocks with slow shock velocities.  To predict the expected line
ratios of characteristic HH emission lines, \citet{Har87}  modelled HH bow
shocks  for a wide range of shock velocities ($V_\rmn{s}$), pre-shock densities
and pre-shock ionization conditions, assuming radiative planar shocks. 
Equilibrium preionization models in the low density regime ($n_\rmn{e}$ $\simeq$
100--1000 cm$^{-3}$) and with low shock velocities ($V_\rmn{s}$ $\leq$~40 \kms\
) give low \nii/H$\alpha$ line ratios ($\sim$ 0.01--0.22, increasing with
$V_\rmn{s}$, with a weak dependence on the electron density) and high
\sii/H$\alpha$ line ratios ($\sim$ 2.5--0.8, decreasing with $V_\rmn{s}$). Thus 
\nii/H$\alpha$ ($\simeq$ 0.2) and \sii/H$\alpha$ ($\simeq$ 0.6) line ratios, 
such as those found in the GH9 knot, are compatible with those obtained in 
shock models with $V_\rmn{s}$ $\simeq$ 30--40~\kms\ occurring in a low density
and partially ionized medium.
Shock models with higher shock velocities ($V_\rmn{s}\geq60$~\kms)
predict \sii/H$\alpha$ line ratios much lower (by a factor of 4), not compatible
with the values derived in HH~262, even considering that the fluxes were
uncorrected for reddening.
Lower \nii/H$\alpha$ line ratios ($\sim$ 0.01) are
found in the planar shock models with  lower velocities   ($V_\rmn{s}$ $\sim$
20~\kms).  The line ratio values derived  in the central knots GH10E-W show a
trend that seems to be compatible, within the errors, with that found in shock
models of very low $V_\rmn{s}$.
The \sii/H$\alpha$ line ratios derived in HH~262 are also well reproduced by the
low-velocity ($V_\rmn{s}\leq40$~\kms) models of \citet{Har94} that include the
effect of the magnetic field, for low $B$ values ($\sim30$ $\mu$G, in
accordance with the current estimates for the Taurus Molecular Cloud,
\citealt{Nak08}), and low ionization fractions ($x_\rmn{e}\leq0.1$).

\subsection{Channel maps}

Since the HH~262 emission spreads over a wide range of velocities,  we will
explore whether the complex, knotty  HH~262 morphology shown in the narrow-band
images is also  found in all the channel maps or, on the contrary, there are
changes in morphology depending on the velocity.

\subsubsection{Spatial distribution of the emission}

\begin{figure}
\centering
\includegraphics[width=\hsize]{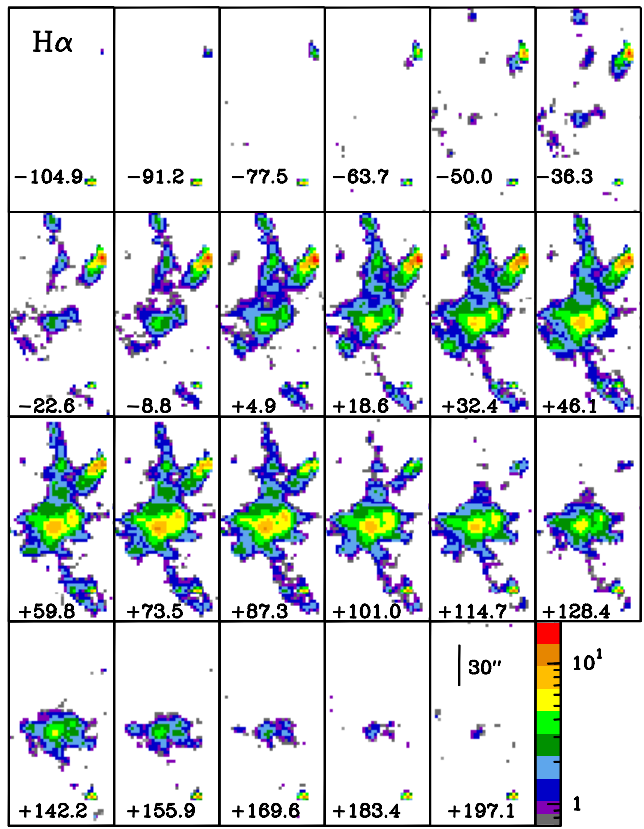}
\caption{
H$\alpha$ channel maps centred on the  \vlsr\  radial velocities 
(km~s$^{-1}$)  labelled in each panel. The spatial scale is given in one of the
panels. Fluxes have been displayed in a logarithmic scale, in  units of
$10^{-16}$ erg~s$^{-1}$~cm$^{-2}$.
\label{cha}}
\end{figure}

\begin{figure}
\centering
\includegraphics[width=0.85\hsize]{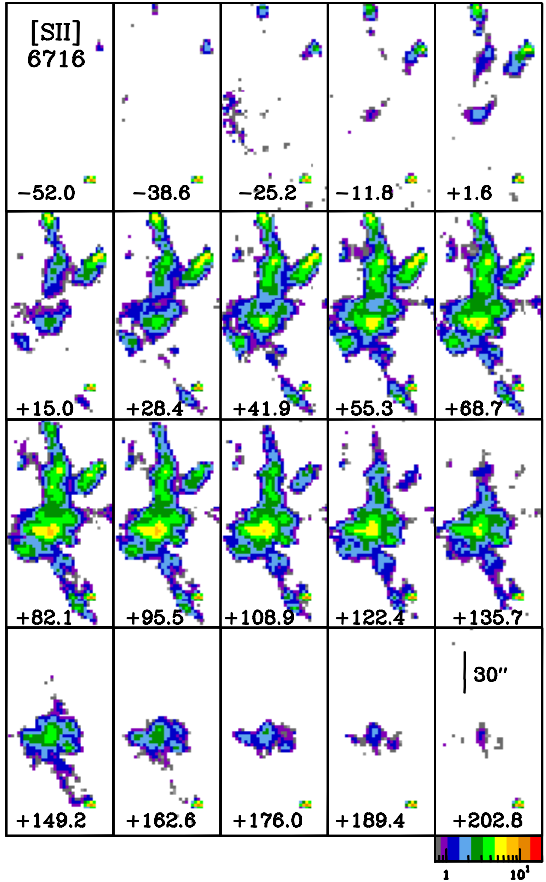}
\caption{
Same as Fig.\ \ref{cha}, but for the \sii\ 6716~\AA\  emission line.
\label{cs1}}
\end{figure}

A more detailed sampling of the gas kinematics was obtained by slicing  the
H$\alpha$ and \sii\ emission of the datacube into a set of velocity channels.
Each slice was made for a constant wavelength bin of 0.3 \AA, corresponding to a
velocity bin of $\sim14$ km~s$^{-1}$ (13.7~km~s$^{-1}$ and 13.4~km~s$^{-1}$ for
H$\alpha$ and \sii\ respectively).  The channel maps obtained for H$\alpha$  and
\sii\ $\lambda$6716 \AA\ are shown in Figs.\ \ref{cha} and \ref{cs1}
respectively.

From these maps, it can be seen that the emission spreads over velocity   ranges
of $\sim290$ km~s$^{-1}$ and $\sim240$ km~s$^{-1}$ wide for  the H$\alpha$ and 
\sii\ $\lambda$6716~\AA\ emissions, respectively. The  HH~262 ''hour-glass''
structure is found for redshifted velocities ranging from  $\sim+5$ to
$\sim+100$~km~s$^{-1}$ in H$\alpha$, and from  $\sim+30$ to $\sim+125$
km~s$^{-1}$ in \sii\ $\lambda$6716~\AA\ .  Furthermore, the channel maps also
show that the northern (K18, K19 and K20) and northwestern (GH9)  knots have
emission at lower velocities as compared  with the other HH~262 knots. These
knots show emission at blueshifted velocities (up to $\sim-50$ km~s$^{-1}$) in
contrast with the southern HH~262 knots (K22), whose emission just begins to
appear at velocities close to the rest velocity of the ambient cloud.  Note in
particular that the bright knot GH9  does not show significant emission for 
velocities higher than $\sim+100$ km~s$^{-1}$, while the emission from the
central (GH10E-W)  and southern  (K22) knots reach velocities up to $\sim$+180
km~s$^{-1}$ and  $\sim$+150 km~s$^{-1}$ respectively. For GH9 and the  northern
knots, the strongest emission corresponds to the velocity channels ranging from
$\sim+19$ to $\sim+45$ km~s$^{-1}$ in  H$\alpha$, and from $\sim +30$ to $\sim
+55$ km~s$^{-1}$ in \sii\ . In contrast, for the central GH10E-W knots,  the
strongest emission appears at higher velocities (in channels ranging from
$\sim+60$ to $\sim+90$ km~s$^{-1}$ in  H$\alpha$, and from $\sim+70$ to
$\sim+110$ km~s$^{-1}$ in \sii\ ).  Thus, we found signatures of deceleration of
the optical  emission as we move to the north, away from the exciting source of
the molecular outflow. Furthermore, an increase of the density in the
northwestern GH9 knot relative to the rest of the HH~262 could also be
contributing to the deceleration of the outflow and thus derive lower velocities
for the GH9 region.

\begin{figure}
\centering
\includegraphics[width=0.8\hsize]{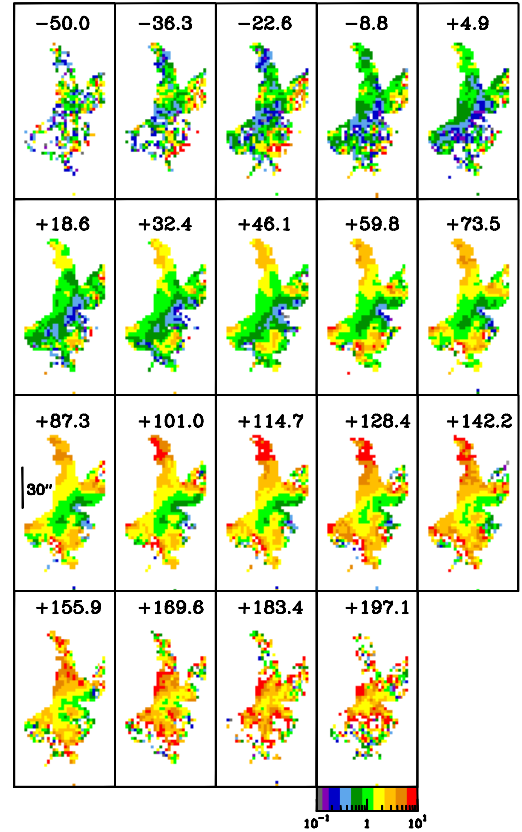}
\caption{
The \sii/H$\alpha$ line-ratio  maps derived from the channel maps at the
velocities labelled in the corresponding panel. 
\label{cden}}
\end{figure}

\subsubsection{Density and excitation conditions}

The behaviour of the density and excitation as a function of velocity was
explored by generating the appropriate line-ratio channel maps. 

We did not find evidence of density variations  as a function of  velocity in
HH~262, as the \sii\ 6716/6731 line-ratio channel maps  show a similar pattern
for all the velocity bins.   The \sii/H$\alpha$ line-ratio channel  maps are
displayed in Fig. \ \ref{cden}. The maps in Fig. \ \ref{cden} show a trend of an
increasing \sii/H$\alpha$ line ratio  (\ie\ a  decreasing excitation) with
increasing velocity. Regarding the spatial distribution, the general trend found
for each velocity channel is a decrease in the line ratios (\ie\ an increase in
the excitation) from east to west across HH~262.   In fact, the behaviour found
in the integrated line-ratio map, where the highest excitation  (lowest
line-ratio values) is found along the western edge of knots GH9 and GH10W, is
also found for each velocity channel.

\subsection{The spectral atlas of HH~262}

A more careful inspection of the HH~262 IFS narrow-band images  (Fig.\
\ref{flux}) shows the very complex morpholohy already found in the CCD images
with higher spatial resolution: most of the named HH~262 knots appear to be
quite
extended  and can be resolved into substructures (\eg\  GH9). As 
already mentioned, IFS data are well suited to characterizing the spectral
properties of such as complex morphology in a more efficient way that by using
long-slit observations.  To take advantadge of this,  we  performed a more
detailed spectral atlas of HH~262. First of all, we obtained a representative 
spectrum for each  knot by co-adding the IFS spectra extracted from the spaxels
inside an appropriate aperture encircling the emission of the
HH~262 knot.  The size of the apertures and the number of spectra that were
co-added to obtain the representative spectrum of the knot, are listed 
in Table 3.

\begin{table}
\begin{minipage}{130mm}
\caption{Apertures defined for the extraction of the knot spectra.
\label {tjcmt}}
\begin{tabular}{@{}llcc@{}}
\hline
\multicolumn{2}{c}{knot}
&Aperture
&Spectra\\
&&(arcsec)$\times$(arcsec)&\\
\hline
K18&&\phantom{1}15\phantom{.}$\times$14\phantom{.5}&46\\
K19&&\phantom{1}8.5$\times$15\phantom{.5}&27\\
K20&&\phantom{1}11\phantom{.}$\times$13\phantom{.5}&30\\
&K20N&\phantom{1}4.5$\times$6.5\phantom{1}&5\\
&K20S&\phantom{1}8.5$\times$6.5\phantom{1}&12\\
GH9&&\phantom{1}13\phantom{.}$\times$34.5&96\\
&GH9N-shell&\phantom{1}2.5$\times$24\phantom{.5}&11\\
&GH9W&\phantom{1}5.5$\times$7.5\phantom{1}&9\\
&GH9E&\phantom{1}8.5$\times$15\phantom{.5}&28\\
GH10W&&\phantom{1}7.5$\times$13\phantom{.5}&20\\
GH10E&&17.5$\times$14\phantom{.5}&51\\
K21&&\phantom{1}7.5$\times$4.5\phantom{1}&8\\
K26&&\phantom{1}9.5$\times$5.5\phantom{1}&10\\
K22&&\phantom{1}11\phantom{.}$\times$13\phantom{.5}&31\\
&K22N&\phantom{1}6.5$\times$7.5\phantom{1}&11\\
&K22C&\phantom{1}6.5$\times$8.5\phantom{1}&12\\
&K22S&\phantom{1}6.5$\times$4.5\phantom{1}&5\\
\hline
\end{tabular}
\end{minipage}
\end{table} 

\begin{figure*}
\centering
\includegraphics[width=\hsize]{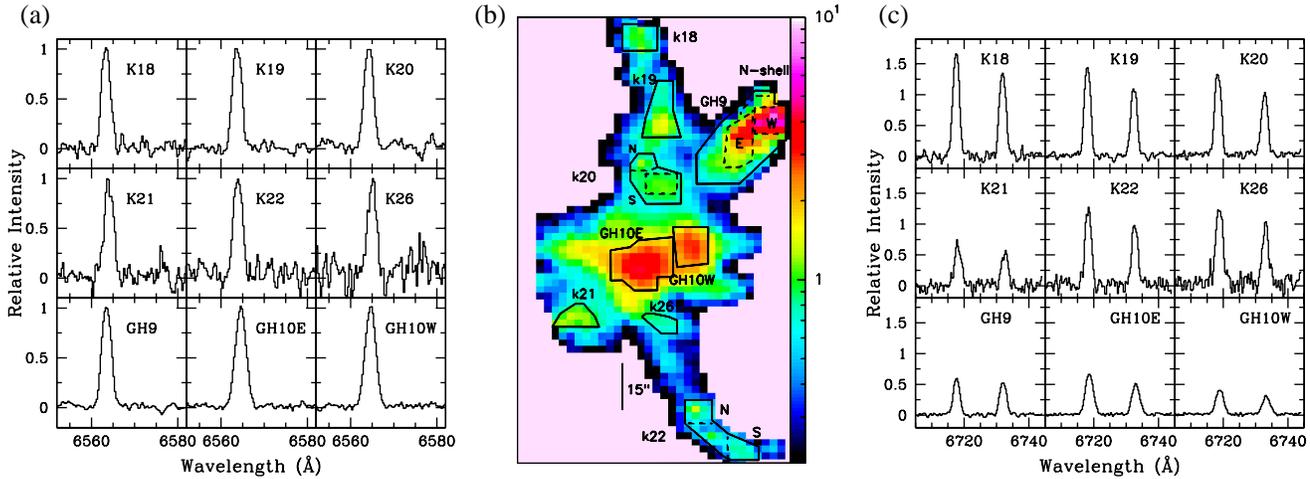}
\caption{Spectra of the HH~262 knots in the spectral windows
including H$\alpha$+\nii\ (a) and \sii\ 6716,6731 emission lines (c). 
Each spectrum has been obtained by co-adding the spectra of the spaxels 
inside an aperture (continuous box)
including the knot emission, and its intensity has been
normalized to the H$\alpha$  peak intensity. The apertures have been marked and
labelled over the H$\alpha$ map of HH~262 (b) and listed in Table 3. Dashed boxes
correspond to the substructures defined inside a given knot.
\label{mapa}}
\end{figure*}

\begin{figure}
\centering
\includegraphics[height=2\hsize]{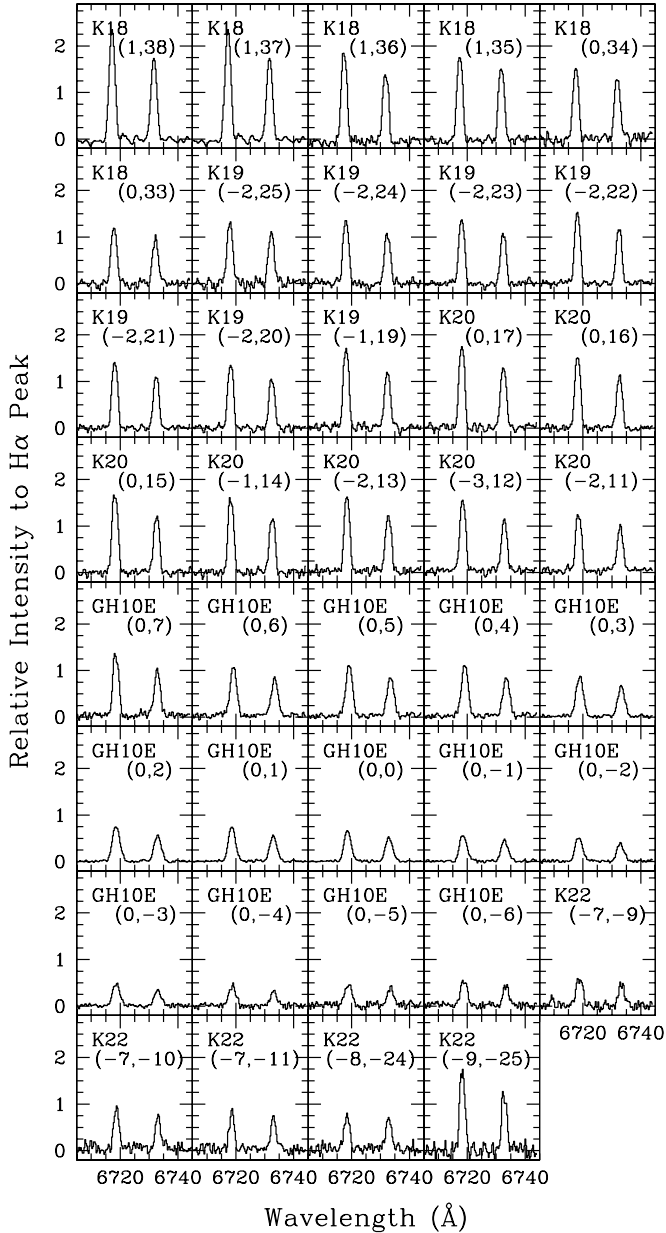}
\caption{HH~262 spectra in the \sii\ spectral window  extracted from
individual spaxels along a longitudinal axis.
Spectra are displayed from north (top left) to south (bottom right). 
Each spectrum corresponds to an aperture of 2.16 $\times$ 2.16 arcsec$^2$ and
has been normalized to the H$\alpha$ peak intensity. The knot where
the spectrum was extracted is indicated in the corresponding panel. 
For each spectrum, the position within HH~262 is indicated by 
the ($\alpha$, $\delta$) 
offset in pixel units (2.16 arcsec~pixel$^{-1}$) from the
reference position. The reference (0,0) position has been set in 
the H$\alpha$ peak position of GH10E.  
\label{cutv}}
\end{figure}

\begin{figure}
\centering
\includegraphics[width=\hsize]{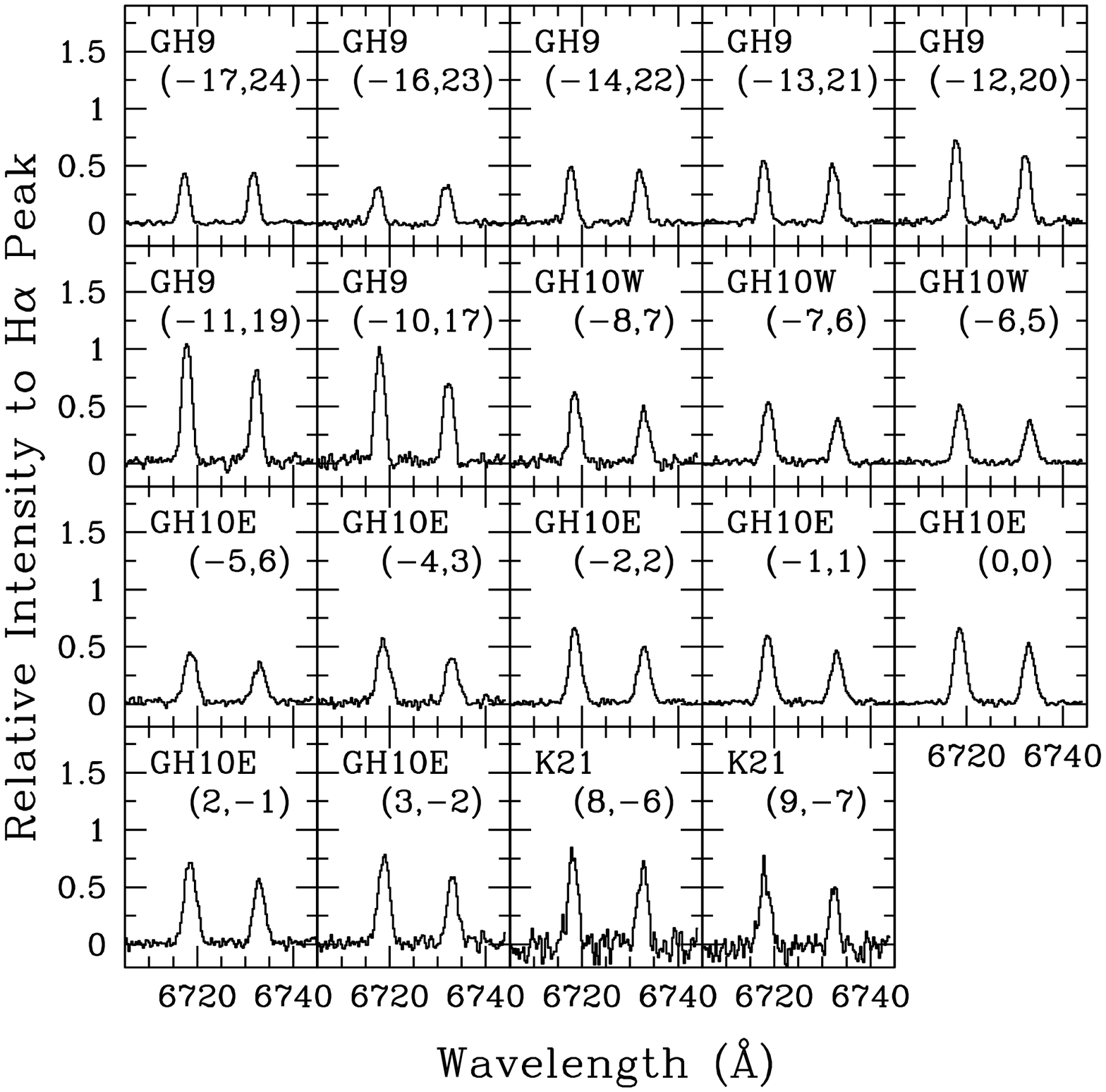}
\caption{Same as Fig.\ \ref{cutv}, but for the spaxels of a 
cut along the transversal, nortwest-southeast 
HH~262 axis (from top left to bottom right).
\label{cuth}}
\end{figure}

Fig.\ \ref{mapa} displays the  spectra of the HH~262 knots, obtained in such a
way,   in the spectral windows covering the H$\alpha$ + \nii\ (left panel) and
\sii\ (right panel) wavelength ranges. In order to make easier comparisons
between the lines in the two spectral windows, the intensities in each of the
spectra were normalized to the corresponding  H$\alpha$ peak intensity.   A
quick inspection of the  spectra in the two spectral  windows suggests that the
excitation conditions   significantly vary from  knot to knot. Note in
particular that the spectrum extracted at the northern knot (K18) presents the
highest \sii/H$\alpha$ line ratio (this indicating the lowest gas excitation). 
Concerning the line profiles,  a more careful inspection of the  spectra
suggests that the line profiles also vary from  knot to knot. In particular, it
can be appreciated that the southern (K21, K22 and K26) and central (GH10W-E)
knots show  more complex, asymmetric line profiles than the northern (K18, K19)
knots for both the H$\alpha$ and \sii\ lines. 

Furthermore, in order to better visualize the spectral variations at a smaller
spatial scale, both along and across HH~262, we extracted spectra  of
individual spaxels for two cuts made along the two main HH~262  axes. One cut
was performed from north to south, intersecting emission from knots K18, K19,
K20, GH10E and K22. The other cut, from northwest to southeast, intersects
emission from  knots GH9, GH10W, GH10E and K21.   The position where the
H$\alpha$ emission has the peak, at the central knot  GH10E, was set as the
origin of positions for referring the location  of a given spectrum.  The
resulting spectra from these two cuts, in the \sii\ window and normalized to
their H$\alpha$ peak intensities, are displayed in Figs.\ \ref{cutv}  and 
\ref{cuth}. The spectra displayed in these  figures suggest there are variations
in the kinematics and physical conditions (excitation and density) along and
across HH~262 even on a short spatial scale ($\le$ 3 arcsec), because we found
appreciable differences, both in the line intensity and in the line profiles
between the spectra extracted at adjacent spaxels.  In particular, the
longitudinal cut (Fig.\ \ref{cutv}) shows the already mentioned increasing in
the gas excitation as we move from the north to the centre of HH~262, suggested
by the decrease along this direction in the \sii\ line intensities relative to
H$\alpha$. This cut also shows that, as a general trend, the line profiles for 
positions north of the central HH~262 knots are more symmetric and single-peaked
as compared with those of the centre and south, where the profiles  appear
wider, suggesting in some cases a contribution from two velocity  components,
partially resolved at our spectral resolution.  Such a regular trend is not
observed in the transversal cut (Fig.\ \ref{cuth}), although the extracted
spectra  reinforced the finding of the short spatial-scale variation conditions
across HH~262.

\subsubsection{Remarks on selected knots}

As has already been mentioned, the H$\alpha$ and \sii\ narrow-band  images of
HH~262 show that the emission from some  knots  appears split into 
subcondensations.  Furthermore, some knots (\eg\ K22 and the central GH10E-W
knots)  show  more complex, asymmetric line profiles, suggesting a contribution
from two velocity components at several positions, while the line profiles are
found to be more symmetric and  single-peaked in the other knots. Thus, in order to 
characterize better the kinematics and physical conditions  in  selected HH~262
knots, we also extracted spectra for each knot subcondensation (see Table 3 and
Fig. \ref{mapa}, central panel for identification)  in addition to the spectrum
extracted for the whole knot emission.

\begin{figure}
\centering
\includegraphics[width=1.0\hsize]{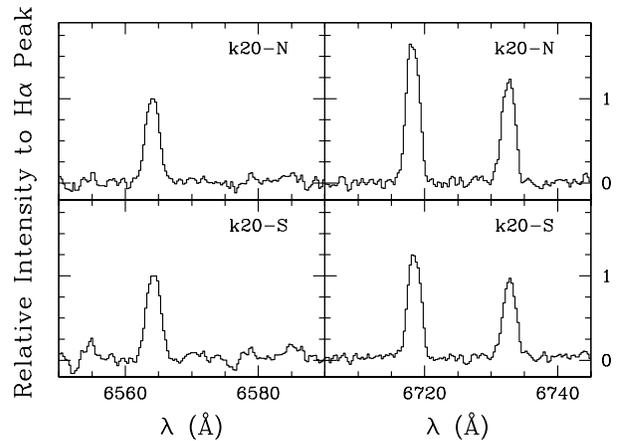}
\caption{Spectra in the H$\alpha$+\nii\ (left) and \sii\ (right) windows 
of the two K20 knot subcondensations obtained by co-adding the IFS spectra
extracted from the spaxels included in the apertures encircling the 
subcondensations. 
In the two
windows, the intensity has been normalized to the corresponding H$\alpha$ peak
intensity.
\label{sk20}}
\end{figure}

\begin{figure}
\centering
\includegraphics[width=1.0\hsize]{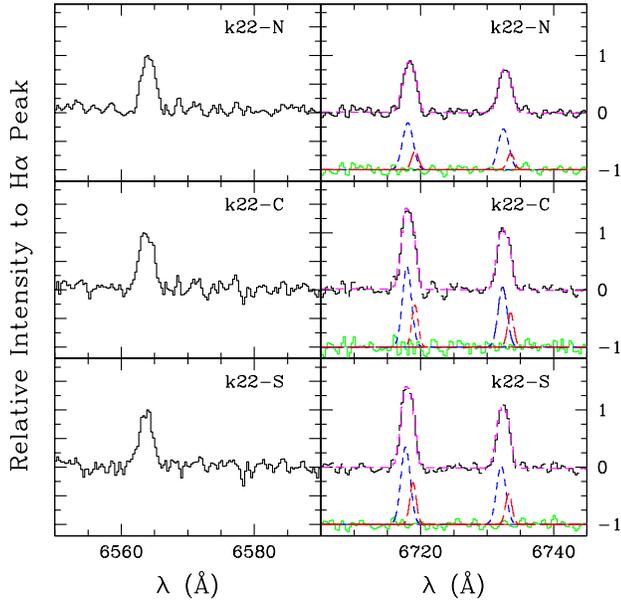}
\caption{Same as Fig.\ref{sk20} but for K22. The fits to the \sii\ line
profiles, obtained from the model, have been drawn in the right panel. The two
velocity components are shown at the bottom. The residual, obtained by
substracting the fit from the observed profile, is plotted in green.   
\label{sk22}}
\end{figure}

\begin{figure}
\centering
\includegraphics[width=1.0\hsize]{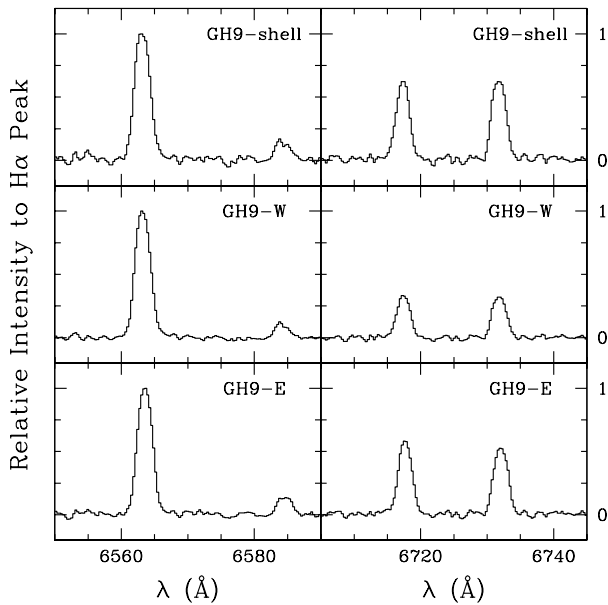}
\caption{Same as Fig. \ref{sk20} but for GH9 knot. 
\label{skgh9}}
\end{figure}

\begin{figure}
\centering
\includegraphics[width=1.0\hsize]{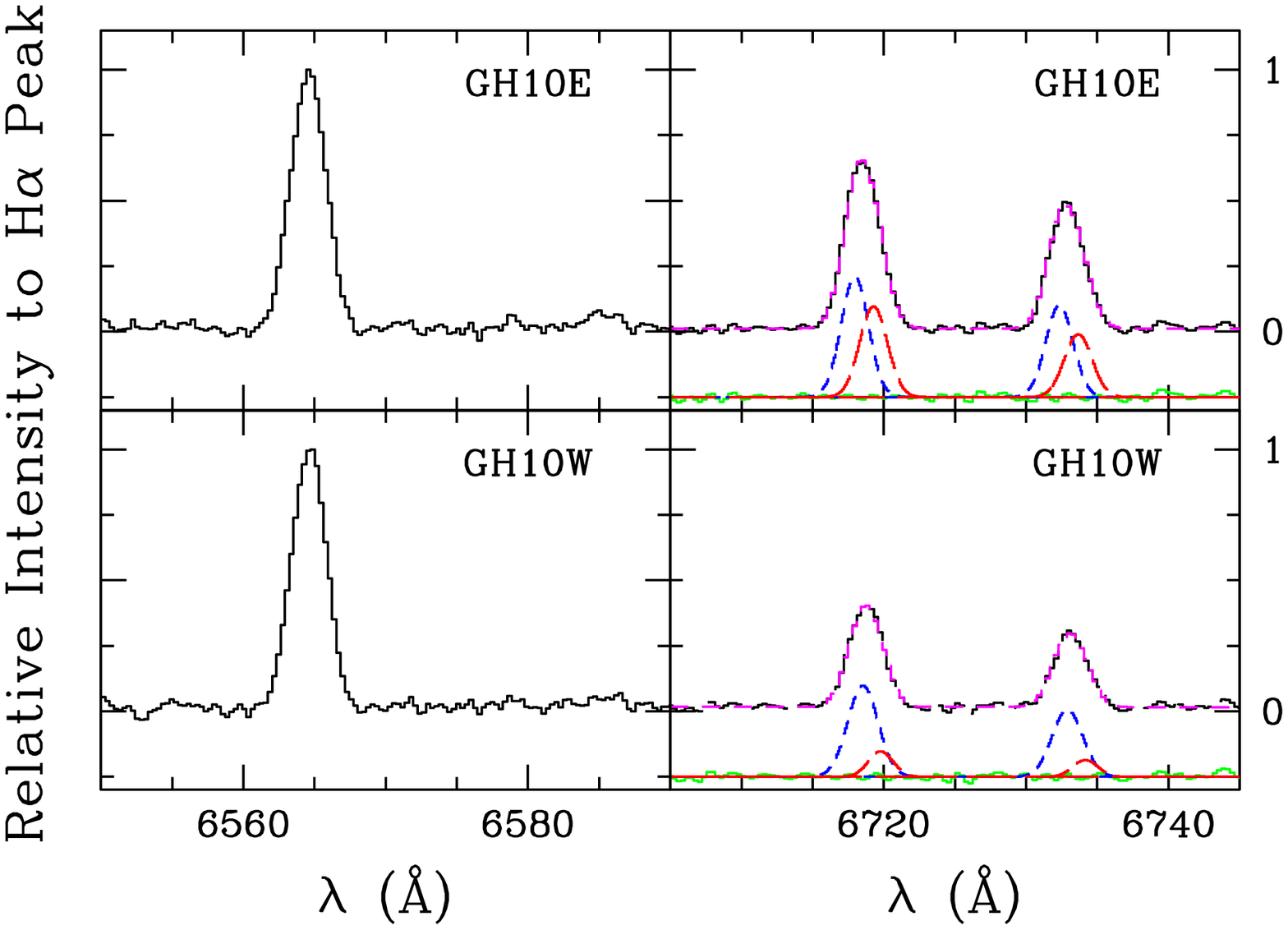}
\caption{Same as Fig. \ref{sk22} but for the GH10E-W knots.
\label{skgh10}}
\end{figure}

\begin{itemize}
\item
{\bf K20:}

Two subcondensations were morphologically defined in knot 20  (K20-N and K20-S),
its peaks being found to be separated by $\sim$ 10 arcsec. We have not found
appreciable kinematic differences between these two subcondensations. As can be
seen in  Fig.\ \ref{sk20}, the H$\alpha$ and \sii\ line  profiles are both 
similar for the two subcondensations as well as the velocities derived from the
centroid of these lines ($\sim$+60~km~s$^{-1}$ and +85~km~s$^{-1}$
respectively). In contrast, a significant change in the \sii/H$\alpha$ line
ratios can be appreciated  when comparing the spectra of the two K20
subcondensations, which  traces an appreciable increase in  the gas excitation
as we move from north to south by only a few arcsec, suggesting a steep gradient
of the gas excitation along  K20.

\item
{\bf GH9:}

The IFS data indicate that the  northwestern arm of HH~262 (\ie\ GH9) seems to
have  different physical conditions compared to the other HH~262 knots. We
extracted the spectra for two subcondensations (GH9W and GH9E) and also for the
lower-brightness shell at the northern GH9 edge (N-Shell) (see  Fig.\ \ref{mapa}
and Table~3 for identification). Their  spectra are shown in  Fig.\ \ref{skgh9}.
Concerning the kinematics, we did  not find appreciable differences among the
GH9 subcondensations. The line profiles appear to be single-peaked  along the
whole of GH9. The velocities derived from the line centroids are similar in the
N-Shell and GH9W,  being slightly higher (by $\sim$ 15 ~km~s$^{-1}$) in GH9E. In
contrast, we can appreciate changes in the physical conditions within GH9 as we
move along the knot. Differences in the excitation conditions (increasing from
N-Shell, GH9E to GH9W) are  present as  the \sii/H$\alpha$ line ratio variations
indicate. Furthermore, the  electron density also varies from one
subcondensation to another.  The highest $n_\rmn{e}$ ($\sim$ 1000 cm$^{-3}$) is
found in the N-Shell, decreasing  to $n_\rmn{e}$ $\sim$ 850 cm$^{-3}$) in GH9W
and  to $n_\rmn{e}$  $\sim$ 600 cm$^{-3}$ in GH9E. Taking into account that
$n_\rmn{e}$ drop  at the centre of HH~262 (\ie\ around knots GH10E-W) until
reaching  values for which the \sii\ line ratio is not
density sensitive in the central region, we can infer that there is a steep density gradient  across 
HH~262 in the  northwest direction.

\item
{\bf K22:}

K22 is one of the HH~262 knots whose extracted spectrum suggests that the line
profiles are double-peaked. Morphologically, the knot appears curved, elongated
and quite extended, but split into subcondensations. We extracted spectra for
each of the three bright K22 subcondensations  (K22-N, K22-C and K22-S, the
separation between the K22-N and K22-S  peaks being $\sim$ 10 arcsec). The
spectra are shown in  Fig.\ \ref{sk22}.  There are appreciable differences in
the \sii/H$\alpha$ line ratios among the K22 subcondensations, this being
indicative of an increase in the gas excitation as we move from south to north
along K22. Thus, a similar gradient in the gas excitation,  but in the opposite
direction to that found in  K20, is detected along K22. The $n_\rmn{e}$ derived
from the \sii\ line ratio increases from south to north along K22, \ie\
following the same trend that the gas excitation.

Concerning the kinematics, both the H$\alpha$ and \sii\ line profiles of the
three subcondensations appear asymmetric and would appear to be  double-peaked. 
We used the \sii\ lines to perform a  decomposition of the line
profiles by applying a model with two Gaussian components.  Gaussian fitting was
perfomed using the {\small DIPSO} package of the {\small STARLINK} astronomical
software. The wavelength difference between the two \sii\ lines was fixed
according to their corresponding atomic parameters. The same FWHM of the
Gaussian in each of the two \sii\ lines was assumed for each component of the
profile. According to the fit parameters, we found two velocity components, both
redshifted and with a  peak separation  of $\sim$  50~km~s$^{-1}$ between them. 
For both components,  the trend found is the peak velocity increases from the
southern to the northern subcondensation. The strongest component has lower
redshifted peak velocities ($V_\rmn{LSR}$ from +55   to +70~km~s$^{-1}$ from
south to north), while the faintest component has  higher redshifted peak
velocities ($V_\rmn{LSR}$ from +105  to +120~km~s$^{-1}$ from south to north).
In Fig.\ \ref{sk22} (right panel), we have also drawn  the fits to the \sii\
line profiles obtained from the  model. The two individual components are shown
at the bottom. Residuals, obtained by subtracting the fit from the  observed
line, are also plotted. 

\item
{\bf GH10:}

Fig.\ \ref{skgh10} displays the spectra of the central GH10E (top) and GH10W
(bottom) knots in the H$\alpha$+\nii\ (left) and \sii\ windows. In  the entire 
area analysed, the spectra extracted at each spaxel of the region   covering 
these two central knots show broad and asymmetric line profiles, suggesting in
most of the positions a double-peaked profile. We performed the line profile
decomposition  analysis described above (K22 subsection) in all the spaxels of
this central region using the \sii\ doublet.  From the fit, we found two
redshifted velocity components spreading over  the whole region,  the component
with the lower velocity being the strongest.  We were unable  to find a clear
spatial separation between these  two components  that would allow the
association  of each of them with a given region of the central knots. However,
a certain trend appears in which the   higher-redshifted component is brighter
towards the east of the GH10E knot, having its  peak intensity somewhat
displaced from the knot centre and  showing in addition a secondary peak
intensity  in a region around  the centre of GH10W.  The lower-redshifted 
velocity component has the peak intensity in the regions located around the
centre  and southeast of GH10E.  

We show in Fig.\ \ref{skgh10} the fits to the \sii\ line profiles obtained from
the model applied to the spectra, obtained as described in \S 3.3, of knots
GH10E and GH10W.  From the fit parameters, we found $V_\rmn{LSR}$ of  +60 and
+120~km~s$^{-1}$ for the two GH10E components and of +85 and +145~km~s$^{-1}$
for these of GH10W knot.

\end{itemize}

\section{Discussion and Conclusions}

We have carried out a study of the kinematics and physical conditions of HH~262
based on IFS observations. From the analysis of these data we find:

\begin{itemize}
 
\item 
HH~262 shows a non-collimated, ''hour-glass'' morphology in
the H$\alpha$ and \sii\ lines. In contrast, the emission from the \nii\ lines 
is  below detection limit except in the northwestern knot GH9. 
This work therefore presents the first
narrow-band \nii\ image currently obtained of HH~262 to show its 
surprisingly different
morphology as compared with the morpholgy found in the other characteristic HH
emission lines. 
\item 
In all the HH~262 knots, radial velocities derived from the line centroids
appear redshifted, relative to the cloud velocity. The highest
velocities and FWHM values are found in the central  knots. 
\item 
A more detailed picture of the kinematics was derived from the channel maps, 
which 
shows  that the emission spreads over a wide range of velocities,  the range
slightly varying for the different knots.  
Furthermore, inspection of the HH~262
extracted spectra shows that there are  broad,  asymmetric line profiles
at the
southern (K22) and central knots (GH10E-W) that are  double-peaked 
at several positions, which would suggest contribution from two velocity 
components. 
The line
profiles appear  more symmetric and single-peaked away from the centre, in the
northern (K18--K20)  and northwestern (GH9) knots.
\item 
Concerning the physical conditions in HH~262, we found variations on a 
spatial scale of $\le$3 arcsec (\ie\ of the order of our spatial resolution). 
As a general trend, the electron densities, derived
from the \sii\ line ratios, are  low. 
In particular, only an upper limit could be set for the brightest,
central HH~262 knots.  Because $n_\rmn{e}$ is referred to the density of the
ionized gas, we propose that the only a small fraction of the 
atomic gas  in HH~262 is  ionized, the remainder 
being mostly neutral gas. It is also plausible that the total
density of the  gas  is lower in this region because the evolved
molecular  outflow has had enough time to clear the medium. Over the entire
velocity range of the emission, the gas
excitation, derived from the \sii/H$\alpha$ line ratios, increases as we move
towards the western region of knot GH10W and to knot GH9  \ie\ moving towards
the edges of HH~262 which
are  closer to the higher extinction region of the cloud. 
The general trend found is thus that
the densest knots have the highest gas
excitation  and the lower centroid velocities.
\end{itemize}

All these  characteristics of HH~262 derived from the IFS data reinforce
 the causal
relationship of the  optical line emission with the two molecular  outflows
(one powered by L1551-IRS5 and the other, by L1551 NE) 
found in the region were
HH~262 is located. In fact, is widely accepted the association between HH~262
and the redshifted lobe of the L1551-IRS5 molecular outflow, first proposed by
\citet{Rod89} and confirmed later on by the Fabry-Perot 
observations \citep{Lop98}.

Detailed PV maps of the L1551-IRS5 outflow performed by \citet{Sto06} from 
high-sensitivity CO observations show a pattern that is characteristic of an
evolved molecular outflow, according to current model simulations of the
entrainment outflow mechanisms  (\citealt{Vel98}, \citealt{Ric00}). In the 
\citet{Sto06} PV CO maps,  the HH objects are found associated with the
molecular emission at the highest velocities, highlighting the 
interaction regions of
the shocks at the head of the shock interface between the  supersonic gas and
the environment. In contrast, the CO emission at lower velocities appears as a
limb-brightened shell-like emission.  Thus, the result derived from the
\citet{Sto06} work support the proposed  scenario, in which the
HH~262 emission traces the walls of a conical  cavity evacuated by an
evolved molecular outflow (L1551-IRS5). In such a scenario, one would 
expect that
the HH~262 radial velocities to decrease (due to projection effects) 
as we move
away from the axis outflow (\ie\ from the central GH10 knots, where we 
found the
highest integrated velocities, to the other knots). It  is also expected
to find double-peaked line profiles along the outflow axis
(corresponding to material  moving in the forward and rear faces of the cone)
and a single-peaked profile away from the axis. Such a trend is
observed in HH~262 from our IFS data, where the lines of the central knots
appear with a complex, asymmetric line profiles, being double-peaked for most
positions, while the line profiles of 
the knots farther away from the outflow axis (\eg\ K18, K19 or GH9) appear 
more symmetric and single-peaked.

In addition to the L1551-IRS5 outflow, the blueshifted lobe of another younger,
dimmer CO outflow, powered by L1551 NE, also encompases HH~262 \citep{Mor06}.
Thus, HH~262 is located in a region where two molecular outflows are interacting
(see  Fig.\ref{diagram1})
and it is expected that this interaction   gives signatures on the jet
turbulence. This interaction could in part be  responsible for the complex
velocity pattern and the wide line profiles found in the optical emission
lines.  In a turbulent jet scenario,  the highest FWHM values are expected  to
be found close to the axis, where there are contributions from gas within a
wider range of velocities because we intersect the entire jet beam, while FWHM
values would be lower at the edge of the jet, away from the axis. This is the
trend found in the IFS HH~262 data for the FWHMs , where the highest values 
correspond to the central GH10 knots and decrease moving towards the edges of
the emission.

The association of HH~262 with an evolved molecular outflow near the edge of a
dark cloud would in part justify the low density values derived for most of the
knots, as the outflow has cleared the medium.  Note however that the   density
of the gas is probably higher than the  density derived from the \sii\ line
ratio, which is an indicator of only the density of the ionized fraction of the
gas rather than of the total density (\citealt{San07}).

Finally, let us discuss the new finding  presented here concerning  the low 
intensity of the \nii\ emission in HH~262. Such a behaviour has  been found in
slow-velocity shocks as those modelled by  \citet{Har87}. In particular, the
\nii/H$\alpha$ and \sii/H$\alpha$ line ratios derived for the knot GH9 are fully
compatible with those derived in shock models with $V_\rmn{s}$ $\simeq$
30--40~\kms\ occurring in a low density, partially ionized medium. Lower
\nii/H$\alpha$ line ratios ($\simeq$ 0.01) are found in models with lower
$V_\rmn{s}$ ($\le$ 20 \kms\ ), thus being compatible, within the errors, with
the marginal detection of \nii\ in the northern knots and with the  line ratios
derived in those knots. The HH~262 line ratios are also compatible with those 
predicted by the shock models of \citet{Har94} for low magnetic field strength
and low ionization fraction values.
Note however that the  morphology of HH~262 is 
too complex and rather different from the morphologies found in most 
 known HH jets.
Thus, it should not be expected that the radiative planar shock 
models of \citet{Har87} and \citet{Har94} 
succeeded in reproducing all the HH~262 properties derived from the current IFS
data. Further IFS observations covering a wavelength range blueward of \nii\ +
H$\alpha$ are needed to search for emission in the \nin\ $\lambda$ 5200 \AA\
line whose strength could justify the weakness of the \nii\ emission found in
HH~262 provided  the nitrogen abundance were not far from the solar value.

\section*{Acknowledgments}
The authors were supported by the Spanish MEC grants 
AYA2005-08523-C03-01 (RE, RL, and AR),
AYA2005-09413-C02-02 (SFS),
AYA2006-13682 (BGL) and 
AYA2005-08013-C03-01 (AR).
In addition we acknowledge 
FEDER funds (RE, RL, and AR), and
the Plan Andaluz de Investigaci\'on of Junta de Andaluc\'{\i}a as 
research group
FQM322 (SFS). 
We appreciate Terry Mahoney's help with the manuscript.

\bsp

\label{lastpage}

\end{document}